\documentclass[14pt]{article}
\usepackage{arxiv}
\usepackage{hyperref}
\usepackage{ulem}
\usepackage{xspace}
\usepackage{graphicx}
\usepackage{fancyhdr}
\usepackage{pslatex}   
\usepackage{xcolor}
\usepackage{bm}
\usepackage{color}
\usepackage{amsmath, amsthm, amssymb} 
\usepackage{amssymb}
\usepackage{enumerate}
\usepackage{enumitem}
\usepackage[symbol]{footmisc}

\usepackage{lineno}
\usepackage[utf8]{inputenc} 
\usepackage[T1]{fontenc}    

\newcommand{\jpsi}{J/\psi}

\newcommand{\xxb}{\Xi^-\overline{\Xi}^+}

\newcommand{\pim}{\pi^{-}}
\newcommand{\lam}{\Lambda}
\newcommand{\lbar}{\overline{\Lambda}}
\newcommand{\xim}{\Xi}
\newcommand{\xibarp}{\overline{\Xi}}

\title{Weak phases and CP-symmetry tests in sequential decays of entangled double-strange
baryons}
\author{
\Large BESIII Collaboration \\
}

\begin{document}
\maketitle
\begin{small}
\begin{center}
M.~Ablikim$^{1}$, M.~N.~Achasov$^{10,c}$, P.~Adlarson$^{67}$, S. ~Ahmed$^{15}$, M.~Albrecht$^{4}$, R.~Aliberti$^{28}$, A.~Amoroso$^{66A,66C}$, M.~R.~An$^{32}$, Q.~An$^{63,49}$, X.~H.~Bai$^{57}$
% ...
\\
\vspace{0.2cm}
\vspace{0.2cm} {\it
$^{1}$ Institute of High Energy Physics, Beijing 100049, People's Republic of China\\
% ...
$^{4}$ Bochum Ruhr-University, D-44780 Bochum, Germany\\
% ...
$^{10}$ G.I. Budker Institute of Nuclear Physics SB RAS (BINP), Novosibirsk 630090, Russia\\
% ...
$^{15}$ Helmholtz Institute Mainz, Johann-Joachim-Becher-Weg 45, D-55099 Mainz, Germany\\
% ...
$^{28}$ Johannes Gutenberg University of Mainz, Johann-Joachim-Becher-Weg 45, D-55099 Mainz, Germany\\
% ...
$^{32}$ Liaoning Normal University, Dalian 116029, People's Republic of China\\
% ...
$^{49}$ State Key Laboratory of Particle Detection and Electronics, Beijing 100049, Hefei 230026, People's Republic of China\\
% ...
$^{57}$ University of Jinan, Jinan 250022, People's Republic of China\\
% ...
$^{63}$ University of Science and Technology of China, Hefei 230026, People's Republic of China\\
% ...
$^{66}$ University of Turin and INFN, (A)University of Turin, I-10125, Turin, Italy; (B)University of Eastern Piedmont, I-15121, Alessandria, Italy; (C)INFN, I-10125, Turin, Italy\\
$^{67}$ Uppsala University, Box 516, SE-75120 Uppsala, Sweden\\
% ...
$^{c}$ Also at the Novosibirsk State University, Novosibirsk, 630090, Russia\\
% ...
}\end{center}

\vspace{0.4cm}
\end{small}
\begin{abstract}  
Using a sample of $1.31\times10^9$ $J/\psi$ events collected with the BESIII detector at the electron--positron collider BEPCII, we analyse the full
$J/\psi\to\Xi^-\overline{\Xi}^+,\ \Xi^-\to\lam \pi^-,\ \lam\to p\pi^-, \overline{\Xi}^+\to\overline{\lam}\pi^+,\ \overline{\lam}\to\overline{p}\pi^+$ decay chain. A new method, exploiting the fact that the $\Xi^-\overline{\Xi}^+$ pair is entangled and sequentially decaying, and where the complete decay chains are reconstructed, is applied for the first time. This enables precision measurements of the decay parameters for the $\Xi^-\to\Lambda\pi^-$ decay ($\alpha_{\Xi}$, $\phi_{\Xi}$) as well as the $\overline{\Xi}^+\to\overline{\Lambda}\pi^+$ decay ($\overline{\alpha}_{\Xi}$, $\overline{\phi}_{\Xi}$). From the decay parameters, two independent CP tests were performed, quantified by the observables $A_{\rm CP}^{\Xi}$ and $\Delta \phi_\Xi$. Our results,
 $A_{\rm CP}^{\Xi}$  = $(6.0\pm13.4\pm5.6)\times10^{-3}$ and $\Delta \phi_\Xi= (-4.8\pm13.7\pm2.9)\times10^{-3}~{\rm rad}$, are consistent with CP symmetry. Furthermore, our method enables a separation of strong and weak $\Xi\to\Lambda\pi$ decay amplitudes. This results in the first direct measurement of the weak phase difference for any baryon decay. The result is found to be $(\xi_{P} - \xi_{S}) =  (1.2\pm3.4\pm0.8)\times10^{-2}$~rad and is one of the most precise tests of CP symmetry for strange baryons. The strong phase difference is measured to be $(\delta_P - \delta_S) = (-4.0\pm3.3\pm1.7)\times10^{-2}$~rad. In addition, we provide an independent measurement of the recently debated $\Lambda$ decay parameter, $\alpha_{\Lambda} = 0.757 \pm 0.011 \pm 0.008 $. The $\Lambda\overline{\Lambda}$ asymmetry is measured to be $A_{\rm CP}^{\Lambda} = (-3.7\pm11.7\pm9.0)\times10^{-3}$.
\end{abstract}

\section{Introduction}

The Standard Model (SM) of particle physics has been proven immensely successful in describing the elementary particles and their interaction. However, there are still questions that the SM cannot answer. One example is why our Universe consists of so much more matter than antimatter. Unless fine-tuned in the Big Bang, the matter abundance must have dynamical origin. This dynamical generation, \textit{Baryogenesis}, is however only possible if certain criteria are fulfilled. One of these is the existence of processes that violate charge conjugation and parity (CP) symmetry \cite{Sakharov:1967dj}. Small violations of CP symmetry are predicted by the SM \cite{Cabibbo:1963yz,Kobayashi:1973fv} and are a well-established phenomenon in weak decays of mesons. However, since these violations are far too small to explain \textit{e.g} the observed matter-antimatter asymmetry of the Universe \cite{Sakharov:1967dj,Bernreuther:2002uj,Canetti:2012zc}, CP tests are considered a promising area to search for beyond-the-Standard-Model (BSM) physics. So far, no CP violating effects beyond the SM have been observed, neither in the meson nor in the baryon sector.

In general, CP symmetry is tested by comparing the decay patterns of a particle to those of its antiparticle. Hadrons decay through an interplay of strong and weak processes, quantified in terms of relative phases between the amplitudes. Many CP-symmetry tests in hadron decays rely on a non-zero strong phase to reveal the signal. This strategy is applied in the determination of the ratio $\epsilon'/\epsilon$, quantifying the difference between the two-pion decay rates of the two weak eigenstates of neutral kaons.
The $\epsilon'/\epsilon$ measurement constitutes the only observation of direct CP violation for strange hadrons \cite{AlaviHarati:1999xp, Fanti:1999nm} and provides the most stringent test of BSM contributions in strange quark systems \cite{Buras:2020xsm}. This strategy, however, comes at a price: it is difficult to disentangle the weak/BSM contributions from the strong processes in a model-independent way. Approaches that do not rely on strong interactions require that the kaon decays into four final state particles \cite{Sehgal:1992wm}. In contrast, baryons allow for this separation through spin measurements. Known examples involving three-body decays are spin correlations and polarisation in nuclear and neutron $\beta$ decays \cite{Gonzalez-Alonso:2018omy}. Sequential two-body decays of entangled multi-strange baryon-antibaryon pairs provide 
another, hitherto unexplored diagnostic tool to 
separate the strong and the weak phases. The approach, outlined in Ref. \cite{Adlarson:2019jtw}, requires simultaneous but independent 
determination of spin-$\frac{1}{2}$ baryon and antibaryon two-body decay parameters. 

\begin{figure}
    \centering
    \includegraphics[width=1.0\textwidth]{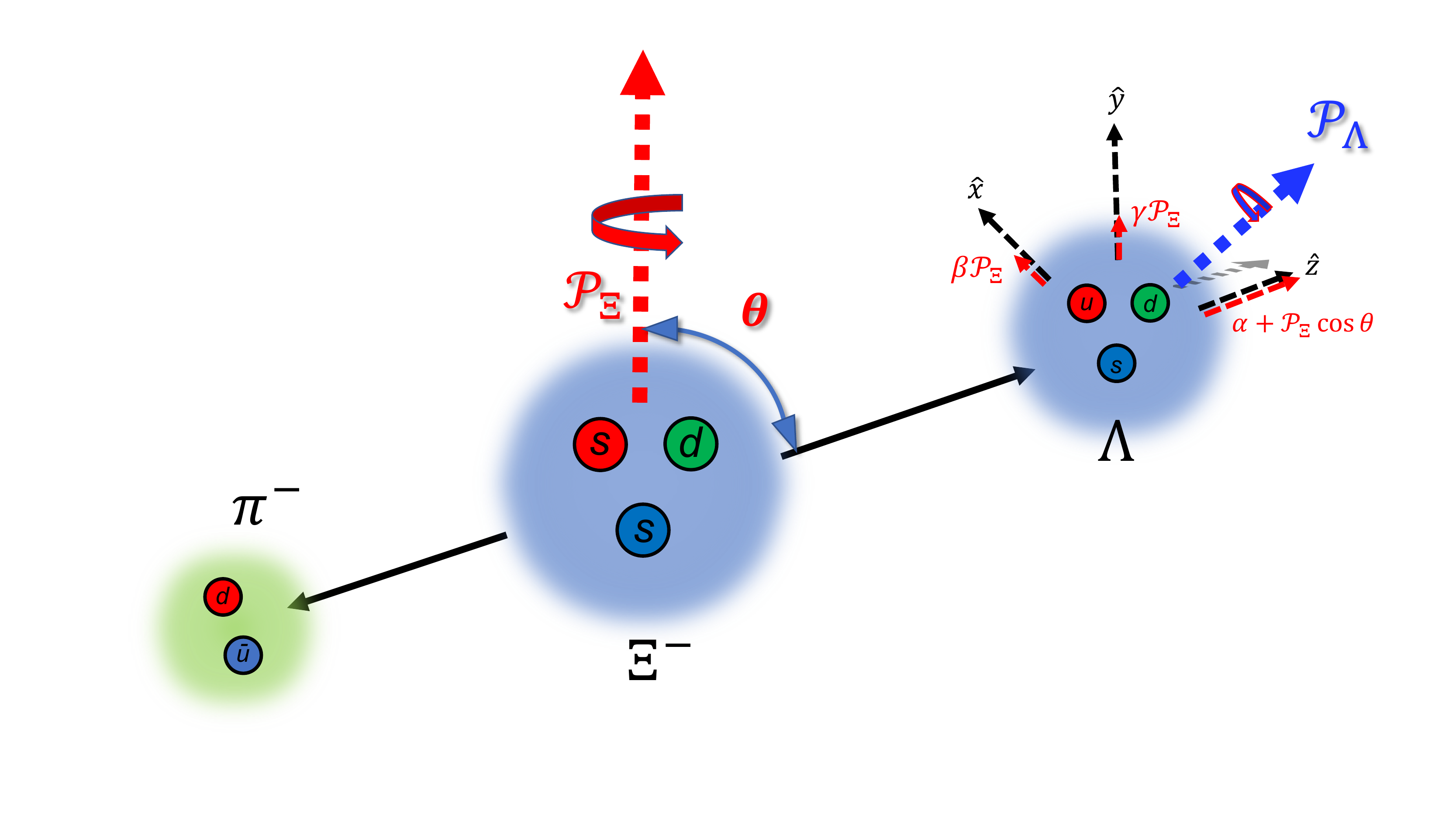}
    \caption{Illustration of the relation between the polarisation vectors of $\Xi$ and $\Lambda$ in terms of the decay parameters $\alpha_\Xi$, $\beta_\Xi$ and $\gamma_\Xi$.}
    \label{fig:decaypars}
\end{figure}

The spin direction of an unstable baryon manifests itself in the momentum direction of its daughter particle, enabling straight-forward experimental access to the spin properties. Spin-$\frac{1}{2}$ baryon decays are described by a parity conserving (P-wave) and a parity violating (S-wave) amplitude, quantified in terms of
the decay parameters $\alpha_Y$, $\beta_Y$ and $\gamma_Y$ \cite{Lee:1957qs}. The $Y$ refers to the decaying mother hyperon (\textit{e.g.} $\Lambda$ or $\Xi^-$). These parameters are constrained by the relation $\alpha_Y^2 + \beta_Y^2 + \gamma_Y^2 = 1$. By defining the parameter $\phi_Y$ according to
\begin{equation}
    \beta_Y = \sqrt{1-\alpha_Y^2}\sin\phi_Y, ~~~\gamma_Y = \sqrt{1-\alpha_Y^2}\cos\phi_Y,
    \label{eq:leeyang}
\end{equation}
the decay is completely described by two independent parameters $\alpha_Y$ and $\phi_Y$. In the standard experimental approach \cite{Cronin:1963zb,Ablikim:2018zay,Dauber:1969hg, Luk:2000zw, Huang:2004jp}, the initial baryon is produced in a well-defined spin polarised state which allows access to the decay parameters through the angular distribution of the final-state particles. For sequentially decaying baryons, \textit{e.g.} the decay of the double-strange $\Xi^-$ baryon into $\Lambda \pi^-$, two effects are possible: 1) a polarised $\Xi^-$ transfers its polarisation $\bm{P}_\Xi$ to the daughter $\Lambda$; 2) a longitudinal component of the daughter $\Lambda$ polarisation is induced by the $\Xi^-$ decay even if the $\Xi^-$ polarisation has no component in this direction. In a reference system with the $\hat{\bf z}$ axis along the $\Lambda$ momentum 
in the $\Xi^-$ rest frame and the $\hat{\bf y}$ axis along ${\bf P}_\Xi\times\hat{\bf z}$, the $\Lambda$ polarisation vector is given by \cite{Lee:1957qs}
\begin{equation}
{\bf P}_\Lambda\cdot\hat{\bf z} =\frac{\alpha_\Xi+{\bf P}_\Xi\cdot\hat{\bf z}}{1+\alpha_\Xi {\bf P}_\Xi\cdot\hat{\bf z}}\ ; ~~~\ {\bf P}_\Lambda\times\hat{\bf z}= { P}_\Xi\sqrt{1-\alpha_\Xi^2}\frac{\sin\phi_\Xi\hat{\bf x} +\cos\phi_\Xi\hat{\bf y}}{1+\alpha_\Xi {\bf P}_\Xi\cdot\hat{\bf z}}\ ,
\label{eq:lambdapol}
\end{equation}
as illustrated in Fig.~\ref{fig:decaypars}. This means that the longitudinal ($\hat{\bf z}$) component depends on $\alpha_{\Xi}$ while the transversal components are rotated by the angle $\phi_{\Xi}$ with respect to the $\Xi^-$ polarisation. 

 The decay parameter $\alpha_\Xi$ appears explicitly in the angular distribution of the direct decay $\Xi^- \to \Lambda \pi^-$, whereas the sequential decay distribution of the daughter $\Lambda$ depends on both $\alpha_{\Lambda}$ and $\phi_\Xi$. CP symmetry implies that the baryon decay parameters $\alpha$ and $\phi$ equal those of the antibaryon $\overline{\alpha}$ and $\overline{\phi}$ but with opposite sign. Hence, CP violation can be quantified in terms of the observables
 \begin{equation}
     A_{\rm CP}^Y = \frac{\alpha_Y+\overline\alpha_Y}{\alpha_Y-\overline\alpha_Y},  ~~~\Delta \phi_{\rm CP} = \frac{\phi_Y+\overline\phi_Y}{2}.
     \label{eq:CPasy}
 \end{equation}
  
CP violation can only be observed if there is interference between CP-even and CP-odd terms in the decay amplitude. Since the decay amplitude for $\Xi^- \to \Lambda \pi^-$ consists of both a P-wave and an S-wave part, the leading-order contribution to the CP asymmetry $A_{\rm CP}^{\Xi}$ can be written as
    \begin{equation}
      A_{\rm CP}^{\Xi} \approx - \tan(\delta_P - \delta_S)\tan(\xi_P - \xi_S),
      \label{eq:acpphase}
  \end{equation}
where $\tan(\delta_P-\delta_S) = \beta/\alpha$ denotes the strong phase difference of the final-state interaction between the $\Lambda$ and $\pi^-$ from the $\Xi^{-}$ decay. CP-violating effects would manifest themselves in a non-zero weak phase difference $\xi_P - \xi_S$ \cite{Donoghue:1986hh, Tandean:2002vy}, an observable that is complementary to the kaon decay parameter $\epsilon^{'}$ \cite{Christenson:1964fg, AlaviHarati:1999xp, Fanti:1999nm} since the latter only involves an $S$-wave.
The strong phase difference can be extracted from the $\phi_\Xi$ parameter, and is found to be small: ($-0.037\pm0.014$) \cite{Huang:2004jp,Zyla:2020zbs}. Hence CP-violating signals in $A_{\rm CP}^\Xi$ are strongly suppressed and difficult to interpret in terms of the weak phase difference.

An independent CP-symmetry test in $\Xi^-\to\Lambda\pi^-$ is provided by determining the value of $\Delta \phi_{\rm CP}$.
At leading order, this observable is related directly to the weak-phase difference
\begin{equation}
 (\xi_P - \xi_S)_{\rm{LO}} = \frac{\beta + \overline{\beta}}{\alpha - \overline{\alpha}}\approx \frac{\sqrt{1-\alpha^2}}{\alpha}\Delta \phi_{\rm CP}  
 \label{eq:betaprime}
\end{equation} 
and can be measured even if $\delta_P = \delta_S$.

This absence of a strong suppression factor therefore improves the sensitivity to CP-violation effects by an order of magnitude with respect to that of the $A_{\rm CP}^{\Xi}$ observable \cite{Donoghue:1985ww, Donoghue:1986hh}. To measure
$\Delta \phi_{\rm CP}$ using the aforementioned standard technique requires beams of polarised 
$\Xi^-$ and $\overline{\Xi}^+$.
In such experiments the precision is limited by the magnitude of the polarisation and 
the accuracy of the polarisation determination, which in turn is sensitive to asymmetries in the production mechanisms \cite{Sozzi:2008zza}. In fact, no experiment with 
a polarised $\overline{\Xi}^+$ has been performed, and the polarisation of the $\Xi^-$ beams were below 5\%  \cite{Huang:2004jp}.
Here, we present an alternative approach where the baryon--antibaryon pair is produced 
in a spin-entangled CP eigenstate and all decay sequences are analysed simultaneously.

Until now, no direct measurements of any of the asymmetries defined in Eq.~\eqref{eq:CPasy} have been performed for the $\Xi^-$ baryon. The HyperCP experiment, designed for the purpose of CP tests in baryon decays, utilised samples of the order $10^7-10^8$ $\Xi^-$ and $\overline{\Xi}^+$ events to determine the products $\alpha_\Xi\alpha_\Lambda$ and $\overline{\alpha}_{\Xi}\overline{\alpha}_{\Lambda}$ \cite{Holmstrom:2004ar}. From these measurements, the sum $A_{\rm CP}^{\Lambda}+A_{\rm CP}^\Xi$ was estimated to be $(0.0 \pm 5.5 \pm 4.4 )\times10^{-4}$. In addition to the aforementioned problem of the smallness of $\phi_{\Xi}$, which limits the sensitivity of $A_{\rm CP}^\Xi$ to CP violation, an observable defined as the sum of asymmetries comes with other drawbacks: should $A_{\rm CP}^\Lambda$ and $A_{\rm CP}^\Xi$ have opposite signs, the sum could be consistent with zero even in the presence of CP-violating effects. A precise interpretation therefore requires an independent measurement of $A_{\rm CP}^\Lambda$ with matching precision. The most precise result so far is a recent BESIII measurement \cite{Ablikim:2018zay} where $A_{\rm CP}^\Lambda$ was found to be $(-6\pm12\pm7)\times10^{-3}$. Furthermore, Ref. \cite{Ablikim:2018zay} revealed a 17\% disagreement with old measurements on the $\alpha_\Lambda$ parameter \cite{Zyla:2020zbs}, a result that rapidly gained some support from a re-analysis of CLAS data \cite{Ireland:2019uja}. Though the CLAS result is in better agreement with BESIII than with the PDG value from 2018 and earlier, there is a discrepancy between the CLAS and BESIII results that needs to be understood.
This is particularly important since many physics quantities from various fields depend on the parameter $\alpha_\Lambda$. Examples include baryon spectroscopy, heavy ion physics and hyperon-related studies at the LHC.

\section{Method}
In this work, a novel method \cite{Perotti:2018wxm,Adlarson:2019jtw} is applied for the first time to study entangled, sequentially decaying baryon-antibaryon pairs in the process $e^+e^- \to J/\psi \to \Xi^{-}\overline{\Xi}^+$. This approach enables a direct measurement of all weak decay parameters of the $\Xi^- \to \Lambda \pi^-, \Lambda \to p \pi^- + c.c.$ decay chain. The production and multi-step decays can be described by nine kinematic variables, here expressed as the helicity angles $\bm{\xi}= (\theta, \theta_{\lam}, \varphi_{\lam}, \theta_{\overline{\lam}}, \varphi_{\overline{\lam}}, \theta_{p}, \varphi_{p}, \theta_{\overline{p}}, \varphi_{\overline{p}})$. The first, $\theta$, is the $\Xi^-$ scattering angle with respect to the $e^+$ beam in the centre-of-momentum (c.m.) system of the reaction. The angles $\theta_{\lam}$ and $\varphi_{\lam}$ ($\theta_{\overline{\lam}}, \varphi_{\overline{\lam}}$) are defined by the $\Lambda$ ($\overline{\Lambda}$) direction in a reference system denoted ${\cal R}_\Xi$ (${\cal R}_{\overline{\Xi}}$), where $\Xi^-$ ($\overline{\Xi}^+$) is at rest and where the $\hat{z}$ axis points in the direction of the $\Xi^-$ ($\overline{\Xi}^+$) in the c.m. system. The $\hat{y}$ axis is normal to the production plane. The angles $\theta_{p}$ and $\varphi_{p}$ ($\theta_{\overline{p}}$ and $\varphi_{\overline{p}}$) give the direction of the proton (antiproton) in the $\Lambda$ ($\overline{\Lambda}$) rest system, denoted ${\cal R}_\Lambda$ (${\cal R}_{\overline{\Lambda}}$), with the $\hat{z}$ axis pointing in the direction of the $\Lambda$ ($\overline{\Lambda}$) 
in the ${\cal R}_\Xi$ (${\cal R}_{\overline{\Xi}}$) system and the $\hat{y}$ axis normal to the plane spanned by the direction of the $\Xi^-$ ($\overline{\Xi}^+$) and the direction of the $\Lambda$ ($\overline{\Lambda}$). The structure of the nine-dimensional angular distribution is determined by eight global (\textit{i.e} independent of the $\Xi^-$ scattering angle) parameters $\bm{\omega}=(\alpha_{\psi},\Delta\Phi,\alpha_{\Xi},\phi_{\Xi},\overline{\alpha}_{\Xi},\overline{\phi}_{\Xi},\alpha_{\lam},\overline{\alpha}_{\lam})$ and can be written in a modular form as \cite{Perotti:2018wxm}:
\begin{equation}
\label{eqn:jointangdis}
    {{\cal{W}}}(\bm{\xi};\bm{\omega}) = \sum_{\mu,\nu = 0}^{3}C_{\mu\nu}\sum_{\mu'\nu' = 0}^{3} a_{\mu\mu'}^{\Xi} a_{\nu\nu'}^{\overline{\Xi}} a_{\mu'0}^{\lam} a_{\nu'0}^{\lbar}.  
\end{equation}
Here $C_{\mu\nu}(\theta;\alpha_{\psi},\Delta\Phi)$ is a $4\!\times\!4$ spin density matrix, defined in the aforementioned reference systems ${\cal R}_{\Xi}$ and  ${\cal R}_{\overline{\Xi}}$, describing the spin configuration of the entangled hyperon--antihyperon pair. The parameters $\alpha_{\psi}$ and $\Delta\Phi$ are related to two production amplitudes, where $\alpha_{\psi}$ parameterises the $\Xi^-$ angular distribution. The $\Delta\Phi$ is the relative phase between the two production amplitudes (in the so-called helicity representation, Ref. \cite{Faldt:2017kgy}) and governs the polarisation $P_y$ of the produced $\Xi^-$ and $\overline{\Xi}^+$ as well as their spin correlations $C_{ij}$. The matrix elements are related to $P_y = P_y(\theta)$ and $C_{ij} = C_{ij}(\theta)$ in the following way:
\begin{equation}
C_{\mu\nu}=(1+\alpha_\psi\cos^2\theta)\left(
\begin{array}{cccc}
1&0&P_y&0\\
0&C_{xx}&0&C_{xz}\\
-P_y&0&C_{yy}&0\\
0&-C_{xz}&0&C_{zz}\\
\end{array}
\right).\label{eqn:cxx}
\end{equation} 
 
The matrices $a^Y_{\mu\nu}$ in Eq.~\eqref{eqn:jointangdis} represent the propagation of the spin density matrices in the sequential decays. The elements of these 4$\times$4 matrices are parameterised in terms of the weak decay parameters $\alpha_Y$ and $\phi_Y$ as well as the helicity angles:  
$a_{\mu\mu'}^{\Xi}(\theta_{\lam},\varphi_{\lam};\ \alpha_{\Xi},\phi_{\Xi})$ in reference system ${\cal R}_{\Xi}$, $a_{\nu\nu'}^{\overline\Xi}(\theta_{\lbar},\varphi_{\lbar};\ \overline{\alpha}_{\Xi},\overline{\phi}_{\Xi})$ in system ${\cal R}_{\overline{\Xi}}$, $a_{\mu'0}^{\lam}(\theta_p,\varphi_p;\ \alpha_{\lam})$ in system ${\cal R}_{\Lambda}$, and $a_{\nu'0}^{\lbar}(\theta_{\overline{p}},\varphi_{\overline{p}};\ \overline\alpha_{\lam})$ in system ${\cal R}_{\overline{\Lambda}}$. The full expressions of $C_{\mu\nu}$ and $a^Y_{\mu\nu}$ are given in Ref.\cite{Perotti:2018wxm}. 

\section{Analysis}
We have carried out our analysis on a data sample of $(1.3106\pm0.0070)\times 10^9$ $J/\psi$ events collected in electron-positron annihilations with the multi-purpose BESIII detector \cite{Ablikim:2009aa}. The $J/\psi$ resonance decays into the $\Xi^{-}\overline{\Xi}^{+}$ final state with a branching fraction \cite{Zyla:2020zbs} of $(9.7 \pm 0.8)\times10^{-4}$. Our method requires exclusively reconstructed $\Xi^{-}\overline{\Xi}^+ \to \Lambda\pi^-\overline{\Lambda}\pi^+ \to p\pi^-\pi^-\overline{p}\pi^+\pi^+$ events. The final state particles are measured in the main drift chamber (MDC), where a superconducting solenoid provides a magnetic field allowing for momentum determination with an accuracy of 0.5\% at 1.0 GeV/$c$. The $\Lambda(\overline{\Lambda})$ candidates are identified by combining $p\pi^-(\overline{p}\pi^+)$ pairs and the $\Xi^{-}(\overline{\Xi}^{+})$ candidates by subsequently combining $\lam\pi^-(\lbar\pi^+)$ pairs. Since it was found that the long-lived $\Xi^-$ and $\overline{\Xi}^+$ can only be reconstructed with sufficient quality if they fulfil $|\cos\theta| < 0.84$, only $\Xi^-$ and $\overline{\Xi}^+$ within this range were considered. After applying all selection criteria, 73 244 $\Xi^{-}\overline{\Xi}^{+}$ event candidates remain in the sample. The number of background events in the signal is estimated to be $187\pm16$. More details are given in the {\bf Appendix}.

For each event, the complete set of the kinematic variables ${\bm{\xi}}$ is calculated from the intermediate and final-state particle momenta. The physical parameters in $\bm{\omega}$ are then determined from ${\bm{\xi}}$ by an unbinned maximum log-likelihood  (MLL) fit where the multidimensional reconstruction efficiency is taken into account. The details of the MLL fit procedure and the systematic uncertainties are described in the {\bf Appendix}.

\section{Results}

\begin{figure}
    \centering
    \includegraphics[width=0.95\textwidth]{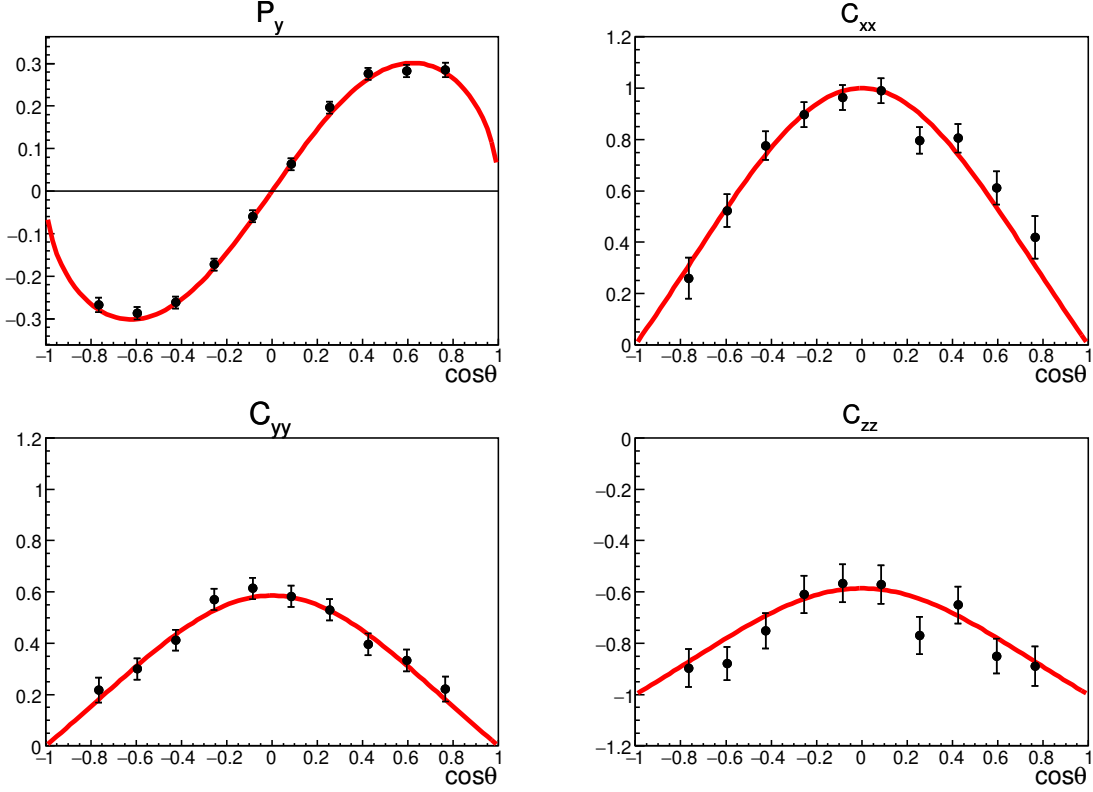}
    \caption{Polarisation and spin correlations and in the $e^+e^-\to\Xi^-\overline\Xi^+$
    reaction. (a) $P_{y}$, (b) $C_{xx}$, (c) $C_{yy}$ and (d) $C_{zz}$ as functions of $\cos\theta$. 
    The data points are determined independently in each bin. The red curves represent the expected angular dependence obtained with the values of $\alpha_{\psi}$ and $\Delta\Phi$ from the global fit. The errors bars indicate the statistical uncertainties.}
    \label{fig:spincorrelations}
\end{figure}

The results of the fit, \textit{i.e.} the weak decay parameters $\Xi^{-}\to\lam\pi^-$ and $\overline{\Xi}^{+}\to\lbar\pi^+$, as well as the production related parameters $\alpha_\psi$ and $\Delta\Phi$, are summarised in Table~\ref{table:main}. 
To illustrate the fit quality, the diagonal spin correlations and the polarisation defined in Eq.~\eqref{eqn:cxx} are shown in Fig.~\ref{fig:spincorrelations}.
The upper left panel of Fig.~\ref{fig:spincorrelations} shows that the $\Xi^-$ baryon is polarised with respect to the normal of the production plane. The maximum polarisation is $\approx30\%$, as shown in the figure.
The data points are determined by independent fits for each $\cos\theta$ bin, without any assumptions on the $\cos\theta$ dependence of $C_{\mu\nu}$. The red curves represent the angular dependence obtained with the parameters $\alpha_\psi$ and $\Delta\Phi$ determined from the global MLL fit. The independently determined data points agree well with the globally fitted curves.

The extracted values of $\alpha_{\Xi}$, $\overline{\alpha}_{\Xi}$, $\phi_{\Xi}$, $\overline{\phi}_{\Xi}$, $\alpha_{\Lambda}$ and $\overline{\alpha}_{\Lambda}$ and their correlations allow for three independent CP symmetry tests. The asymmetry $A^{\Xi}_{\rm CP}$ is measured for the first time and found to be $(6.0\pm13.4\pm5.6)\times10^{-3}$, where the first uncertainty is statistical and the second systematic. The corresponding SM prediction \cite{Tandean:2002vy} is $A_{\rm CP, SM}^{\Xi}=(-0.6\pm1.6)\times10^{-5}$.

The result for the decay parameter $\overline{\phi}_{\Xi}$ is the first measurement of its kind for any weakly decaying antibaryon. By combining this parameter with the corresponding $\phi_{\Xi}$ measurement, the CP asymmetry $\Delta\phi_{\rm CP}^{\Xi}$ can be determined, and is found to be $(-4.8\pm13.7\pm2.9)\times10^{-3}$.
Since this result is consistent with zero, we can improve our knowledge of the value of $\phi_{\Xi}$ by assuming CP symmetry and then calculating the mean value of $\phi_{\Xi}$ and $\overline{\phi}_{\Xi}$. This procedure yields $\left<\phi_{\Xi}\right> = 0.016 \pm 0.014 \pm 0.007$~rad, which differs from the HyperCP measurement, $\phi_{\Xi, \rm HyperCP}=-0.042\pm0.011\pm0.011$~rad, by 2.6 standard deviations \cite{Huang:2004jp}.
It is noteworthy that our method yields a precision in $\left<\phi_{\Xi}\right>$ which is similar to that of the HyperCP result, despite the three orders of magnitude larger data sample of the latter measurement. This demonstrates the intrinsically high sensitivity that can be achieved with entangled baryon-antibaryon pairs. 

The measurement of $\left< \phi_{\Xi} \right>$ allows for a direct determination of the strong phase difference, which is found to be
$(\delta_P - \delta_S) =(-4.0\pm3.3\pm1.7)\times10^{-2}$~rad,
consistent with the SM predictions obtained in the framework of heavy-baryon chiral perturbation theory \cite{Tandean:2002vy} of $(1.9\pm 4.9)\times10^{-2}$~rad but in disagreement with the value $(10.2\pm3.9)\times10^{-2}$ rad extracted from the HyperCP $\phi_{\Xi}$ measurement \cite{Huang:2004jp} using $\alpha_\Xi=-0.376$ from this work.
Since the $(\delta_P - \delta_S)$ value obtained in our analysis is consistent with zero, a calculation of the weak phase difference from Eq.~\eqref{eq:acpphase} is unfeasible. 
Instead, we apply Eq.~\eqref{eq:betaprime}, which yields
$(\xi_{P} - \xi_{S}) =  (1.2\pm3.4\pm0.8)\times10^{-2}$~rad. 
This is one of the most precise tests of the CP symmetry for strange baryons and the first direct measurement of the weak phase difference for any baryon. The corresponding SM prediction \cite{Tandean:2002vy} is $(\xi_P - \xi_S)_{\rm SM} = (1.8\pm1.5)\times10^{-4}$~rad.

Sequential $\Xi^-$ decays also provide an independent measurement of the $\Lambda$ decay parameters $\alpha_{\Lambda}$ and $\alpha_{\overline{\Lambda}}$. Being the lightest baryon with strangeness, many other baryons ($\Sigma^0$, $\Xi^0$, $\Xi^-$, $\Omega$, $\Lambda_c$ etc.) decay with an appreciable fraction into final states containing $\Lambda$. The measurements of spin observables \cite{Adam:2019srw,Becattini:2020ngo,Ablikim:2019vaj} and decay parameters of heavier baryons \cite{Han:2019axh, Blake:2019guk} therefore implicitly depend on $\alpha_{\Lambda}$. Furthermore, since decaying $\Lambda$ and $\overline{\Lambda}$ beams are used for producing polarised proton and antiproton beams \cite{polpbar}, all physics from such experiments rely on a correct determination of $\alpha_\Lambda$. The value of $\alpha_{\Lambda} = 0.757 \pm 0.011 \pm 0.008$ measured in this analysis is in excellent agreement with that obtained from the $J/\psi\to \lam\lbar$ analysis of BESIII \cite{Ablikim:2018zay}, while it disagrees with the result from the re-analysis of CLAS data \cite{Ireland:2019uja}. The precision of our measurement is similar to that of the $J/\psi\to \lam\lbar$ study \cite{Ablikim:2018zay}, despite being based on a six times smaller data sample. The larger sensitivity is primarily explained by the fact that $\alpha_{\Lambda}$ in Eq.~\ref{eqn:jointangdis} appears in a product with the polarisation, which is much larger in the case of $\Lambda$ baryons from $\Xi^-$ decays compared to those directly produced in $J/\psi \to \lam\lbar$.
Furthermore, the multi-step process enhances the angular correlations between the baryons and antibaryons to such an extent that $\alpha_{\Xi}$ and $\alpha_{\Lambda}$ can be measured with the same precision even if the $\Xi^{-}\overline{\Xi}^+$ pair is produced unpolarised \cite{Adlarson:2019jtw}.

\section{Conclusion}
To summarise, this letter presents the first application of a very sensitive test of CP symmetry. This test provides a place to search for BSM physics in strange hadrons that is complementary to $\epsilon'/\epsilon$ measurements in kaon decays. 
When applied to future measurements with larger data sets at BESIII  \cite{Ablikim:2019hff}, the upcoming PANDA experiment at FAIR \cite{panda} and the proposed Super Tau-Charm Factory projects in China and Russia \cite{Bondar:2013cja, Shi:2020nrf}, our method has potential to reach the required precision for CP violating signals, provided such effects exist.

\begin{table}[hbtp]
\centering
\normalsize
\caption[]{{\bf Summary of results.}
The $\jpsi\to\Xi^{-}\overline{\Xi}^+$ angular distribution
  parameter $\alpha_\psi$, the hadronic form factor phase $\Delta\Phi$, the
  decay parameters for $\Xi^{-}\to\lam\pi^-$ ($\alpha_{\Xi},\phi_{\Xi}$),$\overline{\Xi}^{+}\to\lbar\pi^+$ ($\overline{\alpha}_{\Xi},\overline{\phi}_{\Xi}$) $\lam\to p\pim$ $(\alpha_{\lam})$ and $\overline\Lambda\to\overline
  p\pi^+$ $(\overline{\alpha}_{\lam})$; the $CP$ asymmetries $A_{\rm CP}^{\Xi}$, $\Delta\phi_{\rm CP}^{\Xi}$ and $A_{\rm CP}^{\Lambda}$ and the average $\left<\phi_{\Xi}\right>$.  The first and second uncertainties, are statistical and systematic, respectively.}

  \vspace{0.2cm}
  \renewcommand{\arraystretch}{1.3}
\begin{tabular}{lll}
  \hline  \hline
   Parameter  & \multicolumn{1}{c}{This work} & \multicolumn{1}{c}{Previous result} \\
  \hline
  $\alpha_{\psi}$           & $\phantom{-}0.586\pm 0.012 \pm0.010$ &$\phantom{-}0.58\pm0.04 \pm 0.08$ \hfill \cite{Ablikim:2016iym}\\
  $\Delta\Phi$              & $\phantom{-}1.213\pm 0.046 \pm 0.016$~rad &\multicolumn{1}{c}{--} \\

  $\alpha_{\Xi}$           & $-0.376\pm0.007\pm0.003$&$-0.401\pm0.010$\hfill \cite{Zyla:2020zbs}\\
    $\phi_{\Xi}$             & $\phantom{-}0.011\pm0.019\pm0.009$ rad &$-0.037\pm0.014$~rad\hfill \cite{Zyla:2020zbs}\\
  $\overline{\alpha}_{\Xi}$        & $\phantom{-}0.371\pm0.007\pm0.002$ &\multicolumn{1}{c}{--}\\

  $\overline{\phi}_{\Xi}$          & $-0.021\pm0.019\pm0.007$ rad &\multicolumn{1}{c}{--}\\
  $\alpha_{\lam}$           & $\phantom{-}0.757 \pm 0.011 \pm 0.008$ &$\phantom{-}0.750 \pm 0.009\pm 0.004$\hfill \cite{Ablikim:2018zay}\\
  $\overline{\alpha}_{\lam}$          & $-0.763 \pm 0.011 \pm 0.007$ &$-0.758\pm 0.010 \pm 0.007$\hfill\cite{Ablikim:2018zay}\\
    \hline  \hline
    $\xi_P - \xi_S$      & $\phantom{-}(1.2\pm3.4\pm0.8)\times10^{-2}$~rad & \multicolumn{1}{c}{--} \\
    $\delta_P - \delta_S$  & $(-4.0\pm3.3\pm1.7)\times10^{-2}$~rad & $\phantom{-}(10.2\pm3.9)\times10^{-2}$ ~rad\hfill\cite{Huang:2004jp} \\
  \hline  \hline
  $A_{\rm CP}^{\Xi}$ & $\phantom{-}(6.0\pm13.4\pm5.6)\times10^{-3}$ & \multicolumn{1}{c}{--} \\
  $\Delta\phi_{\rm CP}^{\Xi}$ & $(-4.8\pm13.7\pm2.9)\times10^{-3}$ ~rad & \multicolumn{1}{c}{--} \\
  $A_{\rm CP}^{\Lambda}$ & $(-3.7\pm11.7\pm9.0)\times10^{-3}$  &$(-6\pm12\pm7)\times10^{-3}$ \hfill\cite{Ablikim:2018zay} \\
  \hline  \hline
  $\left<\phi_{\Xi}\right>$            & $\phantom{-}0.016 \pm 0.014 \pm 0.007$~rad & \\
  \hline  \hline
\end{tabular}
  \vspace{0.3cm}
 \label{table:main}
\end{table}

\noindent{\Large\bf Acknowledgements}

The BESIII collaboration thanks the staff of BEPCII and the IHEP computing center for their strong support. This work is supported in part by National Key Research and Development Program of China under Contracts Nos. 2020YFA0406300, 2020YFA0406400; National Natural Science Foundation of China (NSFC) under Contracts Nos. 11625523, 11635010, 11735014, 11822506, 11835012, 11905236, 11935015, 11935016, 11935018, 12075107, 11961141012; the Chinese Academy of Sciences (CAS) Large-Scale Scientific Facility Program; Joint Large-Scale Scientific Facility Funds of the NSFC and CAS under Contracts Nos. U1732263, U1832207; CAS Key Research Program of Frontier Sciences under Contracts Nos. QYZDJ-SSW-SLH003, QYZDJ-SSW-SLH040; 100 Talents Program of CAS; CAS President’s International Fellowship Initiative (PIFI) programme; INPAC and Shanghai Key Laboratory for Particle Physics and Cosmology; ERC under Contract No. 758462; European Union Horizon 2020 research and innovation programme under Contract No. Marie Sklodowska-Curie grant agreement No 894790; German Research Foundation DFG under Contracts Nos. 443159800, Collaborative Research Center CRC 1044, FOR 2359, GRK 214; Istituto Nazionale di Fisica Nucleare, Italy; Ministry of Development of Turkey under Contract No. DPT2006K-120470; National Science and Technology fund; Olle Engkvist Foundation under Contract No. 200-0605; STFC (United Kingdom); The Knut and Alice Wallenberg Foundation (Sweden) under Contract No. 2016.0157; The Royal Society, UK under Contracts Nos. DH140054, DH160214; The Swedish Research Council; U. S. Department of Energy under Contracts Nos. DE-FG02-05ER41374, DE-SC-0012069; National Science Centre (Poland) under Contract No. 2019/35/O/ST2/02907.

\noindent{\large\bf Author Contributions}

\noindent All authors have contributed to the publication, being variously
involved in the design and the construction of the detectors, in
writing software, calibrating sub-systems, operating the detectors and
acquiring data and finally analysing the processed data.

\noindent{\large\bf Additional Information}

\noindent{\bf Correspondence and requests for materials} should be addressed to\\ besiii-publications@ihep.ac.cn. \\ 
This pre-print is based on the style template found at  \href{https://github.com/kourgeorge/arxiv-style}{github.com/kourgeorge/arxiv-style}.

\noindent{\large\bf Data Availability Statement}

\noindent The data that support the plots within this paper and other findings of this study are available from the corresponding author upon reasonable request.
\clearpage

\renewcommand\thefigure{(supplementary) \arabic{figure}}    
\renewcommand\thetable{(supplementary) \arabic{table}}    
\setcounter{figure}{0} 
\setcounter{table}{0} 

\section*{Appendix}
\label{sec:Appendix}
\subsection*{Monte Carlo simulation\label{DataMC}}
For the selection and optimization of the final event sample, estimation of background sources as well as normalization for the fit method, Monte Carlo simulations have been used. The simulation of the BESIII detector is implemented in the simulation software \verb!GEANT4!\cite{geant4,Allison:2006ve}. \verb!GEANT4! takes into account the propagation of the particles in the magnetic field and particle interactions with the detector material. The simulation output is digitized, converting energy loss to pulse heights and points in space to channels. In this way the Monte Carlo digitized data have the same format as the experimental data. The production of the $J/\psi$ is simulated by the Monte Carlo event generator \verb!KKMC!\cite{Jadach:1999vf}. Particle decays are simulated using the package \verb!BesEvtGen!\cite{Lange:2001uf,Ping:2008zz}, where the properties of mass, branching ratios and decay lengths come from the world averaged values\cite{Zyla:2020zbs}. 
We find that while the mass of the $\Lambda$ in our data agrees with the established value, that of the $\Xi$ is   $95$~keV/$c^{2}$ 
above the central value of the world average\cite{Zyla:2020zbs}, $m_{\Xi, \rm PDG}$= $1321.71\pm 0.07$~MeV/$c^2$. Hence, we have adjusted the value in the simulation accordingly.

The signal channels used for optimization and consistency checks are implemented with the helicity formalism and with parameter values in close proximity to the results presented in Table~\ref{table:main}.

\subsection*{Selection Criteria}
\begin{figure}
    \centering
    \includegraphics[width=1.0\textwidth]{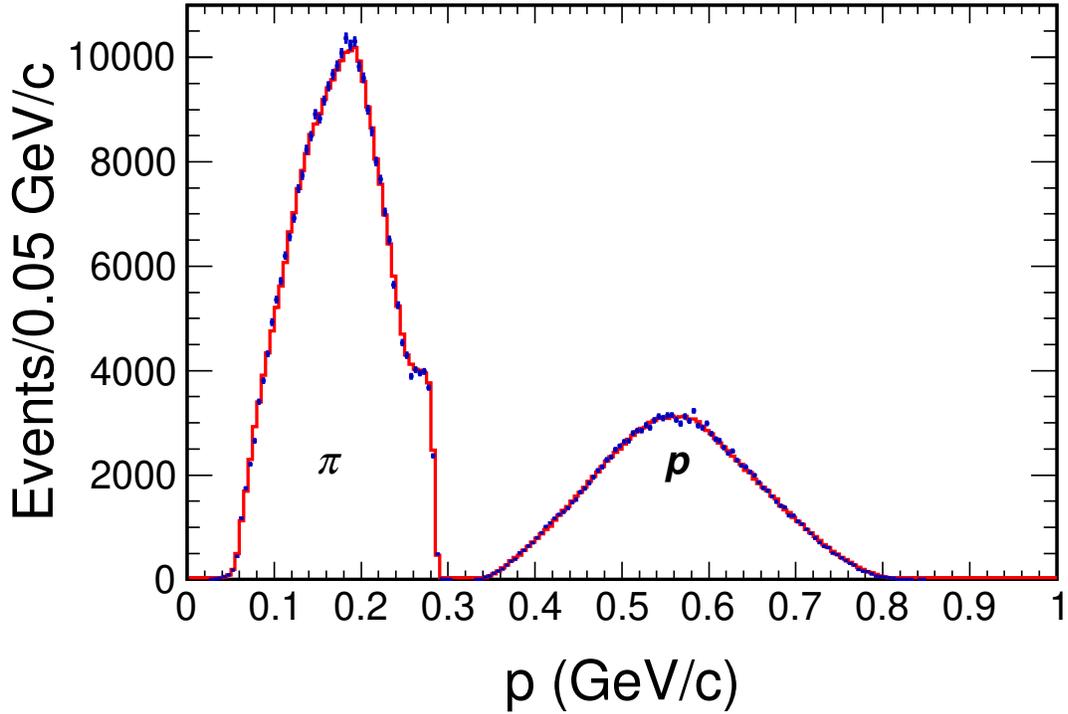}
    \caption{Final-state particle momenta of pions and protons for experimental data (blue crosses) and simulated data (red, solid line).}
    \label{fig:momentumpionsprotons}
\end{figure}

\begin{figure}
    \centering
    \includegraphics[width=0.495\textwidth]{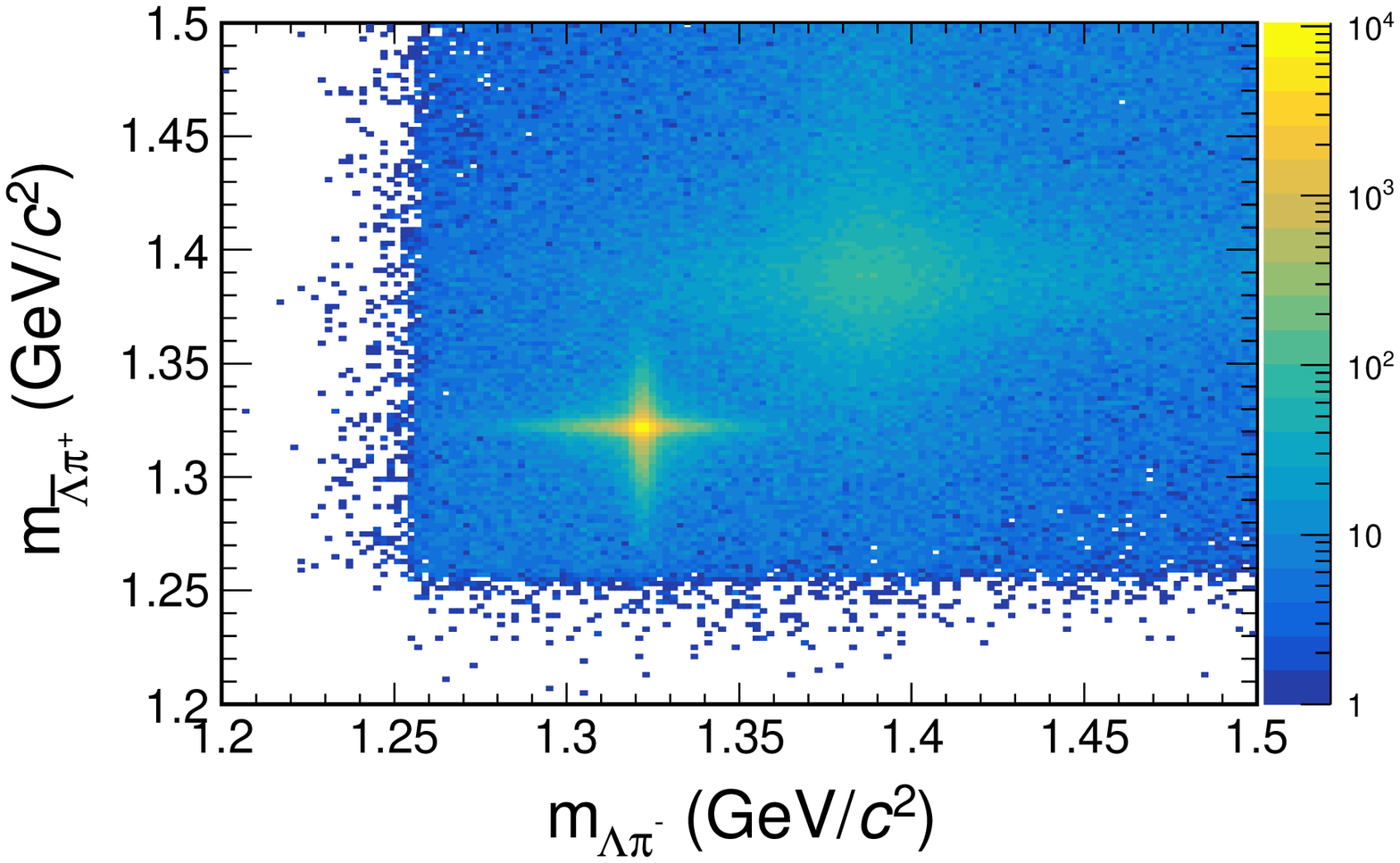}
    \includegraphics[width=0.495\textwidth]{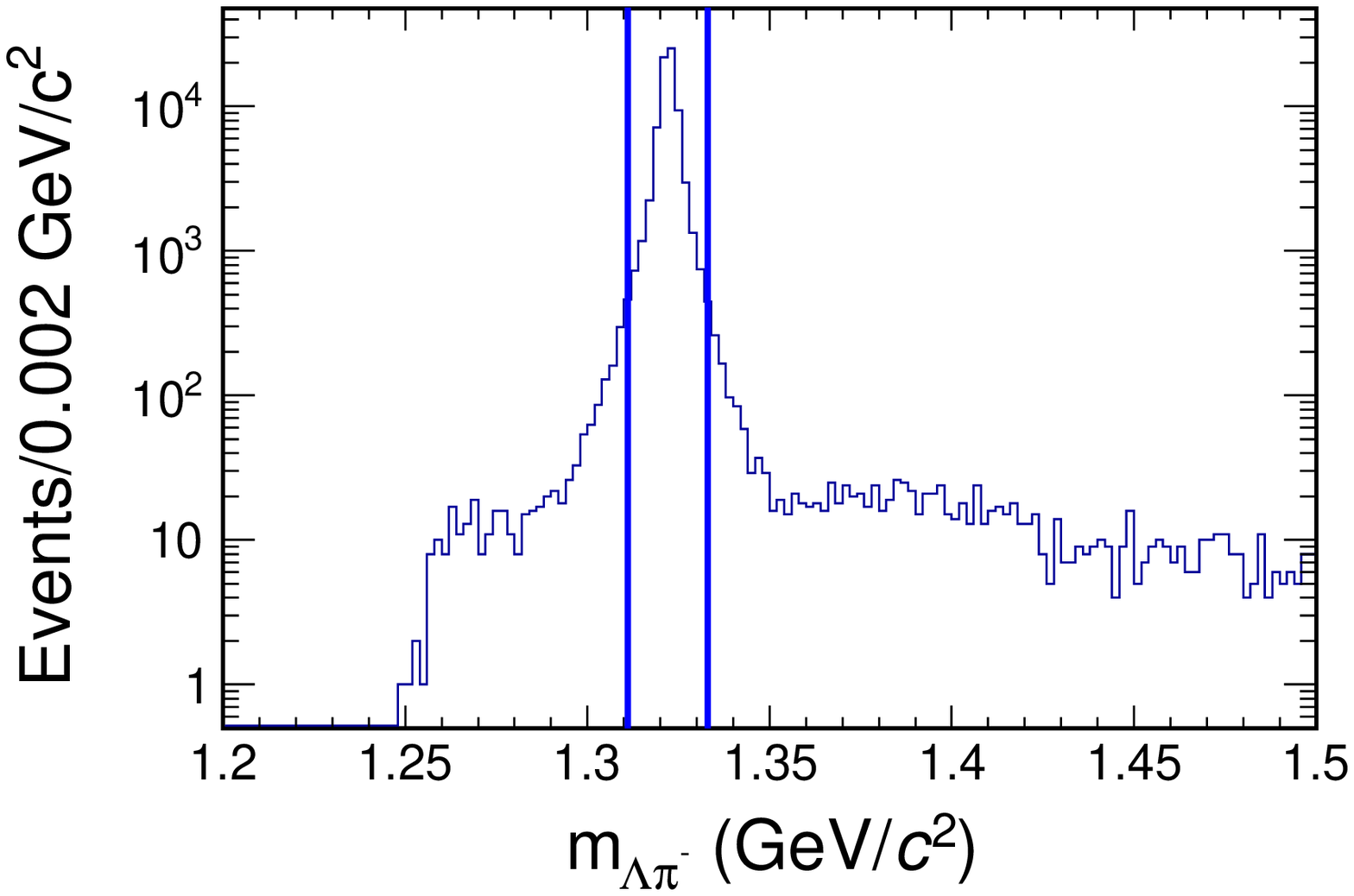}
    \caption{(left) Distribution of the invariant masses $m_{\Lambda
        \pi^-}$ versus $m_{\lbar \pi^+}$. The other structure seen is
      from the reaction
      $J/\psi\to\Sigma(1385)^-\overline{\Sigma}(1385)^+$. (right) The
      $m_{\Lambda \pi}$ distribution for the experimental data sample
      before the final event selection criteria on $m_{\Lambda
        \pi}$ has been applied. The final requirement selects the events between the two lines.}
    \label{fig:mxivmxibarmcfinal}
\end{figure}

The data were accumulated during two run periods, in 2009 and 2012, where the later set is approximately five times larger compared to the earlier.
For the analysis all charged final state particles have to be reconstructed. The Main Drift Chamber (MDC) of the BESIII experimental setup is used for reconstructing the charged-particle tracks. At least three positively and three negatively charged tracks are required, each track fulfilling the condition that
$|\cos\theta|<0.93$, where $\theta$ is the polar angle with
respect to the positron beam direction. The momentum distributions of protons and pions from the signal process are well separated and do not overlap, as shown in Figure \ref{fig:momentumpionsprotons}. Therefore a simple momentum criterion suffices for particle identification:  $p_{pr} > 0.32$~GeV/$c$ and $p_{\pi} < 0.30$~GeV/$c$ for protons and pions, respectively. Only events with at least one proton, one anti-proton, two negatively and two positively charged pions are saved for further analysis. Each $\Xi$ decay chain is reconstructed separately, and is here described  for the sequence $\Xi
^-\to\lam\pi^- \to p \pi^-\pi^-$. To find the correct $\Xi^-$ and $\Lambda$ particles all proton and $\pi^-$ candidates are combined together. The $\Lambda$ and $\Xi^-$ particles are reconstructed through vertex fits by first combining the $p\pi^{-}_{i}$ pair to form a $\Lambda$ and then the $\Lambda \pi^{-}_{j}$ ($i\neq j$) pair to form a $\Xi^-$. The fits take into account the non-zero flight paths of the hyperons which can give rise to different production and decay points. All vertex fits must converge and the combination which minimises $((m_{p \pi\pi}-m_{\Xi})^{2} + (m_{p \pi}-m_{\Lambda})^{2})^{1/2}$, where $m_{\Xi}$ and $m_{\Lambda}$ are the nominal masses and $m_{p\pi\pi}$ ($m_{p\pi}$) is the mass of the candidate $\Xi$ ($\Lambda$), is retained for further analysis. The same procedure is performed for the $\xibarp$ decay chain. For each decay chain the probability that the pions from the $\Xi\to\Lambda\pi$ and $\Lambda\to p \pi$ decays is wrongly assigned is found to be 0.51\% and 0.49\% for $\pi^{+}$ and $\pi^{-}$, respectively, which is negligible for the analysis. 
The $m_{\Lambda \pi
^-}$ versus $m_{\lbar \pi
^+}$ scatter plot is shown in the left panel of Figure \ref{fig:mxivmxibarmcfinal}. A four-constraint kinematic fit requiring energy and momentum conservation ($4C$) is imposed on the $e^{+}e^{-}\to J/\psi\to\xxb$ system, and only events where $\chi^2_{ 4C}<100$ are retained for further analysis. The kinematic fit is effective for removing the background processes $e^{+}e^{-}\to J/\psi\to \gamma \eta_c \to \gamma \xxb$ and $e^{+}e^{-}\to J/\psi\to \Xi(1530)^- \xibarp \to \pi^0 \xxb$ (and its charge conjugate), which have the same charged final state topology as the signal channel, but contain extra neutral particles.

The invariant masses of the $p\pi^-$ and $\overline{p}\pi^+$ pairs are also required to fulfil $|m_{p \pi}- m_{\lam, \rm peak}|<11.5$~MeV/$c^2$, where $m_{\lam, \rm peak}$ is the peak position of the $\Lambda$ mass distribution. A similar mass window criterion, optimised to remove the broad-resonance $\Sigma^{-}(1385)\overline{\Sigma}^{+}(1385)$ background contribution, is imposed on the $\Xi$ particle, $|m_{\Lambda \pi}- m_{\Xi, \rm peak}|<11.0$~MeV/$c^2$.

The decay length is defined as the distance between the point of origin and the decay position of the decaying $\Lambda$ or $\Xi$ particle.
If the hyperon momentum points oppositely to the direction from the collision to the decay point, then the decay length becomes negative in the vertex-fit algorithm. These events are removed from the  sample. 

When comparing experimental data with Monte Carlo simulations there are differences seen for large polar angles. This discrepancy is seen to induce systematic biases  on the parameter values. The differences can, however, be isolated and removed by restricting the sample to events where $|\cos\theta_{\xim, \rm c.m.}|<0.84$. Here $\theta_{\xim, \rm c.m.}$ is the polar angle of the reconstructed $\Xi$ particle in the center-of-mass frame. After applying all selection criteria given above, 73 244 $\Xi^{-}\overline{\Xi}^{+}$ event candidates remain in the final sample.

\subsection*{The maximum log-likelihood fit procedure} 
The global fit is performed on the data through the joint angular distribution. For $N$ events the likelihood function is given by
\begin{equation}
{\cal{L}} (\boldsymbol{\xi}_1, \boldsymbol{\xi}_2,..., \boldsymbol{\xi}_N ;\bm{\omega}) =
\prod_{i=1}^N{\cal{P}} (\boldsymbol{\xi}_i;\bm{\omega})= 
\prod_{i=1}^N\frac{{\cal{W}}(\boldsymbol{\xi}_i;\bm{\omega})\epsilon({\boldsymbol{\xi}_i})}{{\cal{N}}(\bm{\omega})},
\label{eqn:linlik}
\end{equation}
where  $\epsilon({\boldsymbol{\xi}})$ is the efficiency, ${\cal{W}}(\boldsymbol{\xi};\bm{\omega})$ is the weight as specified in Eqn.~(\ref{eqn:jointangdis}), and the normalization factor ${\cal N}(\bm{\omega})=\int {\cal W}(\boldsymbol{\xi};\bm{\omega})\epsilon({\boldsymbol{\xi}}){\rm d}\boldsymbol{\xi}$. The normalisation factor is approximated as  ${\cal N}(\bm{\omega})\approx \frac{1}{M}\sum_{j=1}^M {\cal W}(\boldsymbol{\xi}_j;\bm{\omega})$, using $M$ Monte Carlo events $\boldsymbol{\xi}_j$ generated uniformly over phase space, propagated
through the detector and reconstructed in the same way as data. $M$ is chosen to be significantly larger than the number of events in data $N$; our results exploit a simulation sample where $M/N \sim 35$. By taking the natural logarithm of the joint probability density, the efficiency function can be separated and removed as it only affects the overall log-likelihood normalization and is not dependent on the parameters in $\bm{\omega}$. To determine the parameters, the MINUIT package from the
CERN library is used\cite{James:1975dr}. The minimised function is given by ${\cal S} = -\ln({\cal{L}})$.
The operational conditions were slightly different for the 2009 and 2012 data sets, most notably in the nominal value of the magnetic field. For this reason the likelihoods are constructed separately for the two different run periods.

The results of the simultaneous fit are shown in Table \ref{table:main}. Those results that depend on combinations of decay parameters account for the correlations between the parameters. The correlation coefficients between the decay parameters are given in Table \ref{table:maincorrelations}. Assuming that CP-symmetry is conserved we find $\left<\alpha_{\Lambda}\right>=\phantom{-}0.760 \pm 0.006 \pm 0.003$ and $\left<\alpha_{\Xi}\right> = -0.373 \pm 0.005 \pm 0.002$, where the latter result is in disagreement with the current standard value $\alpha_{\Xi} = -0.401\pm0.010$\cite{Zyla:2020zbs}. The parameter $\alpha_{\Xi}$ has previously only been measured indirectly via the product $\alpha_{\Xi}\alpha_{\Lambda}$ and the assumed value of $\alpha_{\Lambda}$. The current standard value of  $\alpha_{\Lambda}= 0.732\pm 0.014$ is an average based on the two incompatible results of BESIII and the re-analysed CLAS data Ref.\cite{Ablikim:2018zay,Ireland:2019uja}, and in disagreement with the value found in this analysis. In contrast, our measured value for the product $\left<\alpha_{\Xi}\right>\left<\alpha_{\Lambda}\right>=-0.284 \pm 0.004 \pm 0.002$ is compatible with the world average $\alpha_{\Xi}\alpha_{\Lambda} = -0.294\pm0.005$\cite{Zyla:2020zbs}.

{
\begin{table}
\renewcommand{\arraystretch}{1.2}
\normalsize
\caption{Correlation coefficients for the production and asymmetry decay parameters.}
\label{table:maincorrelations}
\begin{center}
\begin{tabular}{| c | c | c | c | c | c | c | c | c | c |}
\hline\hline
&  $\alpha_{\psi}$ & $\Delta\Phi$ & $\alpha_{\Xi}$ & $\phi$ & $\alpha_{\lam}$ & $\overline{\alpha}_{\Xi}$ & $\overline{\alpha}_{\lam}$ & $\overline{\phi}_{\Xi}$ \\ \hline
$\alpha_{\psi}$  & 1.0 & 0.414 & -0.008 & -0.006 & -0.107 & 0.014 & 0.120 & 0.003\\
$\Delta\Phi$  &  & 1.0 & -0.016 & 0.016 & -0.133 & 0.008 &0.138 & -0.029 \\
$\alpha_{\Xi}$  &  &  & 1.0 & -0.000 & 0.280 & 0.024 & 0.071 & 0.010 \\
$\phi_{\Xi}$  &  &  &  & 1.0 & -0.002 & -0.010 & -0.010 & 0.013\\
$\alpha_{\lam}$  &  &  &  &  & 1.0 & 0.070 & 0.401 & 0.014 \\
$\overline{\alpha}_{\Xi}$  &  &  &  &  &  & 1.0 & 0.269 & 0.001 \\
$\overline{\alpha}_{\lam}$  &  &  &  &  &  &  & 1.0 & 0.006\\
$\overline{\phi}_{\Xi}$  &  &  &  &  &  &  &  & 1.0\\
\hline\hline
\end{tabular}
\end{center}
\end{table}
}
\subsection*{Systematic uncertainties}
\label{sec:syserr}
The systematic uncertainties are assigned by performing studies related to the kinematic fit, the $\Lambda$ and $\Xi$ mass window requirements, the $\Lambda$ and $\Xi$ decay length selection, and a combined test on the $\xim\xibarp$ fit reconstruction with the $p,\pi$ MDC track reconstruction efficiency. Searches of systematic effects are tested by varying the criteria above and below the main selection. For each test, $i$, the parameter values are re-obtained, $\bm{\omega}_{\rm sys, i}$ and the changes evaluated compared to the central values, $\Delta_i= |\bm{\omega} - \bm{\omega}_{\rm sys, i}|$. Also calculated are the uncorrelated uncertainties $\sigma_{\rm uc, i} = \sqrt{|\sigma_{\omega}^2 - \sigma_{\omega,\rm sys, i}^{2}|}$, where $\sigma_{\omega}$
and $\sigma_{\omega,\rm sys, i}$ correspond to the fit uncertainties of the main and systematic test results, respectively. If the ratio $ \Delta_i/\sigma_{\rm uc, i}$ shows a trending behaviour and larger than two this is attributed to a systematic effect\cite{Barlow:2002yb, OB13}. 
For each systematic effect the corresponding uncertainty is evaluated. The assigned systematic uncertainties are given in Tables \ref{table:summarytablesyst1}, \ref{table:summarytablesyst2} and \ref{table:summarytablesyst3}, where the individual systematic uncertainties are summed in quadrature.

\begin{enumerate}[]

\item \textbf{Estimator.}
To test if the method produces systematically biased results a large Monte Carlo data sample is produced with production and decay distributions corresponding to those of the fit results to the data sample  ($\sim$10 times the experimental data). The simulated data are divided into subsamples with equal number of events as the experimental sample, and run through the fit procedure. The obtained fit parameters are found to be consistent within one standard deviation of the generated parameter values and hence no bias is detected.

\item \textbf{Kinematic fit.}
The systematic differences from the kinematic fit are tested by varying the kinematic fit $\chi^2$ value from 40 to 200, with an increment of 20 in each step. Significant effects are seen for the parameters $\Delta\Phi$, $\phi$ and $\overline{\phi}$ when $\chi^{2}_{4C}>100$. For $\chi^{2}_{4C}<100$ systematic deviations occur for $\alpha_{\Xi}$ and $\overline{\alpha}_{\lam}$. The difference in track resolution between data and Monte Carlo is the likely cause for these change in the parameter values. The systematic uncertainty is assigned to be the average difference of the main result to a lower and upper limit, determined to be at $\chi^2_{4C} = 60$ and $200$, respectively.

\item \textbf{$\Lambda$ and $\Xi$ mass window selection.}
Possible systematic effects due to the $\Lambda$, $\lbar$, $\Xi^-$ and $\overline{\Xi}^+$ mass windows are investigate by varying the selection criteria between 2-30 MeV/$c^2$ and 2-20 MeV/$c^2$ for the $\Lambda/\lbar$ and $\Xi^-/\overline{\Xi}^+$ candidates, respectively. For the $\Lambda$ selection systematic deviations are seen for decreasing mass windows. 
The uncertainty is assigned to be the difference of the nominal result to the result when $95\%$ of the events are included, at $|m_{p \pi}- m_{\lam, \rm peak}|<6.9$~MeV/$c^2$.
For the $\Xi^-$ and $\overline{\Xi}^+$ mass windows significant effects are seen for the parameters  $\alpha_{J/\psi}$, $\alpha_{\lam}$ and $\phi_{\Xi}$. The systematic uncertainties for these parameters are assigned to be the difference of the main result and the results obtained one standard deviation lower than to the main selection window, estimated from the $m_{\Lambda\pi}$ line shape uncertainty. 

\item \textbf{$\Lambda$ and $\Xi$ decay length.}
Possible systematic effects related to the $\Lambda$ and $\Xi$ life times are studied by varying the decay length selection criteria for the $\Lambda$ and $\Xi$ candidates. For $\Xi$ no strong trending behaviors are seen, but for $\Lambda$ a dependence is seen for the asymmetry parameters $\alpha_{\lam}$ and $\overline{\alpha}_{\lam}$ which is accounted for in the final systematic uncertainty.

\item \textbf{The combined efficiency of $\Xi^-\overline{\Xi}^+$ reconstruction and $p,\pi^-$ tracking.}
For the study of systematic effects related to the tracking  and the $\Lambda$ and $\Xi$ reconstruction it is assumed that the combined efficiency for
proton, anti-proton and $\pi^\pm$ only depends on the polar angle $\cos\theta$ and the transverse momentum, $p_{T}$. To study the tracking efficiency the fitted probability density function is modified by allowing for arbitrary efficiency corrections as a function of $\cos\theta$ and $p_{T}$ for each particle type in an iterative procedure. The correction procedure is repeated until the maximum log likelihood is stable within $\ln(2)$ between two successive iterations. The difference between the fit results with and without the tracking correction is assigned as the systematic uncertainty.

\item \textbf{$\cos\theta_{\Xi,\rm c.m.}$ scattering angle.}
From comparing data to MC simulation a discrepancy is seen for charged tracks with polar angles $|\cos\theta|>0.84$. The discrepancy is also seen to have a notable effect on some of the the decay parameters. The effect can be isolated by only removing the events where $|\cos\theta_{\Xi, \rm c.m.}|>0.84$. Although the observed data-simulation differences are removed by requiring that $|\cos\theta_{\Xi, \rm c.m.}|<0.84$, residual systematic effects are observed for $\left<\alpha_{\Xi}\right>$ and $\left<\alpha_{\Xi}\right>\left<\alpha_{\Lambda}\right>$, which are included in the systematic uncertainty.

\item \textbf{Data set consistency.}
When comparing the statistically independent results of the 2009 and 2012 data sets, all parameters are found to agree within two standard deviations. As there is no evidence of systematic bias, no uncertainty is assigned associated with possible data set differences.

\end{enumerate}

{
\begin{table}
\renewcommand{\arraystretch}{1.2}
\normalsize
\caption{Contributing systematic uncertainties, and the sum in quadrature. First row: statistical uncertainty as reference. The uncertainties of $\Delta\Phi$ and $\phi$ are given in radians. All values multiplied by a factor $10^2$.}
\label{table:summarytablesyst1}
\begin{center}
\begin{tabular}{|c|c| c | c | c | c | c | c | c | c |}
\hline\hline
$\times 10^{2}$&  $\alpha_{\psi}$ & $\Delta\Phi$ & $\alpha_{\xim}$ & $\overline{\alpha}_{\Xi}$ & $\alpha_{\lam}$ & $\overline{\alpha}_{\lam}$ & $\phi_{\Xi}$& $\overline{\phi}_{\Xi}$\\ \hline
Statistical   & 1.2 & 4.6 & 0.70 & 0.70 & 1.05 & 1.06 & 1.91 & 1.93 \\ \hline
Kin. fit & 0.36 & 1.5 & 0.18 & 0.17  & 0.21 & 0.43 & 0.77 & 0.44 \\
mass win $\Lambda$ & 0.44 & 0.44 & 0.07 & 0.02  & 0.56 & 0.33 & 0.17 & 0.46 \\
mass win $\Xi$ & 0.25 & - & - & - & 0.36 & - & 0.46 & - \\
dec. length $\lam$ & - & - & - & - & 0.30 & 0.40 & - & - \\
Track. eff. & 0.80  & 0.41   & 0.27    & 0.05    & 0.21  & 0.14  & 0.16 & 0.16\\ \hline
Sum syst. & 1.0 & 1.6 & 0.33 & 0.18 & 0.79 & 0.69 & 0.93 & 0.66 \\
\hline\hline
\end{tabular}
\end{center}
\end{table}
}

\begin{table}
\renewcommand{\arraystretch}{1.2}
\normalsize
\caption{Contributing systematic uncertainties to CP tests, and the sum in quadrature. First row: statistical uncertainty as reference. All values multiplied by a factor $10^2$.}
\label{table:summarytablesyst2}
\begin{center}
\begin{tabular}{| c | c | c | c | c | c |}
\hline\hline
$\times10^{2}$&  $A_{\Lambda, \rm CP}$ & $ A_{\Xi, \rm CP}$ & $ \Delta\phi_{\rm CP}$ (rad) & $\delta_P - \delta_S$ (rad) & $\zeta_P - \zeta_S$ (rad)  \\ \hline
Statistical  &1.17 & 1.34 & 1.37 & 3.3 & 3.4 \\ \hline
Kin. fit & 0.32  & 0.47 & 0.16 & 1.3 & 0.4 \\
mass win. $\Lambda$ & 0.59  & 0.07 & 0.14 & 0.8 & 0.4  \\
mass win. $\Xi$ & 0.38  & - & 0.20 & 0.7 & 0.5  \\
dec. length $\lam$ & 0.46  & - & - & - & -  \\
Track. eff. & 0.05  & 0.29 & 0.003 & 0.4 & $2\cdot10^{-3}$\\  \hline 
Sum syst. & 0.90 & 0.56 & 0.29 & 1.7 & 0.75 \\
\hline\hline
\end{tabular}
\end{center}
\end{table}

\begin{table}
\renewcommand{\arraystretch}{1.2}
\normalsize
\caption{Contributing systematic uncertainties to average values of decay parameters, and the sum in quadrature. First row: statistical uncertainty as reference. All values multiplied by a factor $10^2$.}
\label{table:summarytablesyst3}
\begin{center}
\begin{tabular}{| c | c | c | c | c |}
\hline\hline
$\times10^{2}$ & $\left<\alpha_{\Xi}\right>$ & $\left<\alpha_{\Lambda}\right>$ & $\left<\phi\right>$ (rad) & $\left<\alpha_{\Xi}\right>\cdot\left<\alpha_{\lam}\right>$  \\ \hline
Statistical  & 0.49 & 0.58 &  1.35 & 0.38 \\ \hline
Kin. fit & 0.09 & 0.19 & 0.54 & 0.02  \\
mass win. $\Lambda$ & 0.05  & 0.12 & 0.31 & $<10^{-2}$  \\
mass win. $\Xi$ & -  & 0.07 & 0.26 & 0.04  \\
$\cos\theta_{\Xi, \rm c.m.}$ & 0.12  & - & - & 0.13  \\
Track. eff.  & 0.16  & 0.17 & 0.16 & 0.06  \\ \hline
Sum syst. & 0.22 & 0.29 & 0.69 & 0.21\\ \hline

\hline\hline
\end{tabular}
\end{center}
\end{table}

\end{document}